\documentclass[prl,aps,showpacs,amsmath,amssymb,superscriptaddress,preprint]{revtex4}
\usepackage{graphicx}
\usepackage{dcolumn}
\usepackage{bm}
\usepackage{subfigure}
\usepackage{setspace}

\begin{document}

\title{Compression of Atomic Phase Space Using an Asymmetric \emph {One-Way} Barrier}

\author{M.G. Raizen}
\author{A.M. Dudarev}
\affiliation{Department of Physics, The University of Texas,
Austin, Texas 78712-1081} \affiliation{Center for Nonlinear
Dynamics, The University of Texas, Austin, Texas 78712-1081}
\author{Qian Niu}
\affiliation{Department of Physics, The University of Texas,
Austin, Texas 78712-1081}
\author{N. J. Fisch}
\affiliation{Princeton Plasma Physics Laboratory, Princeton University,
Princeton, NJ 08543}

\date{\today}

\begin{abstract}
We show how to construct asymmetric optical barriers for atoms.  These barriers can be used to compress phase space of a sample by creating a confined region in space where atoms can accumulate with heating at the single photon recoil level.  We illustrate our method with a simple two-level model and then show how it can be applied to more realistic multi-level atoms. 
\end{abstract}


\maketitle

Laser cooling of atoms relies on the Doppler shift to preferentially scatter near-resonant light when atoms are moving towards the beam, with multiple scatterings required to achieve substantial cooling~\cite{metcalf99}.  Although this approach has been very successful, the process requires a cycling transition, which has limited the applicability of laser cooling to a small set of atoms in the periodic table.  Further cooling below the single photon recoil limit was made possible by creating dark states in momentum space using quantum interference~\cite{aspect88} or stimulated Raman transitions~\cite{kasevich92}. Dark states in position space have been based on creating selective regions where laser cooling turns off due to optical pumping to a dark state~\cite{ketterle93,morigi98}. More recently collective-emission-induced cooling was demonstrated using an optical cavity~\cite{chan03}. Compression of phase space beyond laser cooling has been accomplished by evaporation.

We consider in this Letter a different approach to compression of phase space which utilizes an asymmetric optical barrier that confines atoms in one state but not another.  Spontaneous emission is used only as an irreversible way to transfer atoms from one state to another when they are in the trap.  The laser then reduces the entropy of the atomic cloud by increasing the density with a minimal increase in the kinetic energy.  
\begin{figure}[b]
\includegraphics[width=2.55in]{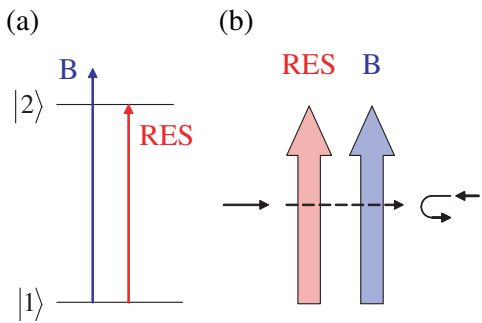}
\caption{\label{fig:1} The first scheme for uni-directional wall. Beam B blue detuned from the resonance creates repulsive potential for atoms in state $\left| 1 \right\rangle $. Beam RES is tuned to atomic resonance.}
\end{figure}

The original motivation for this work came from plasma physics where it was shown that a ponderomotive potential in the radio frequency regime in a magnetic field could be made as an efficient asymmetric barrier for electrons or ions, thereby driving electrical currents with little power dissipation~\cite{fisch03}. Thus, the question to ask is: can we construct in the optical regime a barrier that transmits atoms coming from one side but reflects atoms coming from the other side? 

\begin{figure}[t]
\includegraphics[width=2.55in]{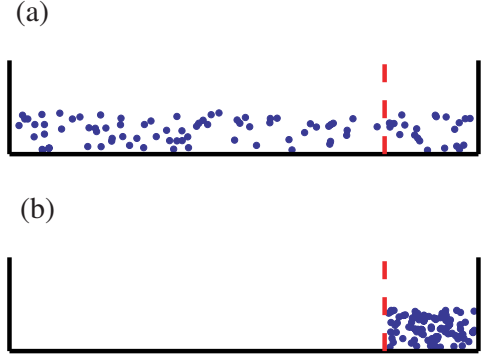}
\caption{\label{fig:2} Illustration of the phase compression process. As the uni-directional wall is placed inside of a box, atoms are accumulated in the smaller part, thus increasing the density. Kinetic energy increase is due to photon recoil as  atoms decay to ground state.}
\end{figure}

To answer this question, we first construct a simple model.  Consider a two-level atom with ground state $\left| 1 \right\rangle $ and an excited state $\left| 2 \right\rangle $ that decays spontaneously back to the ground state with a lifetime $\tau$.  One laser beam, denoted B is tuned to the blue of the atomic transition, while another beam, denoted  RES, is tuned exactly on resonance, as shown in Fig.~\ref{fig:1}(a).  We construct a barrier as shown in Fig.~\ref{fig:1}(b); On the left side is a focused RES sheet, and to the right of that a focused B sheet.  An atom impinging from the right will encounter the B sheet which is a repulsive barrier and it will be reflected back.  In contrast, an atom impinging on the barrier from the left will first be promoted to the excited state $\left| 2 \right\rangle$ with some probability.  It then encounters the barrier which is attractive for that state, so it goes through (neglecting quantum reflection).  We must assume that the spontaneous lifetime is longer than the transit time of the atom through the barrier, and that the atom decays to the ground state after crossing the barrier.  Clearly, this wall reflects atoms from the right and transmits them from the left.  How can such a barrier be used to compress phase space?  Consider a 1D (one-dimensional) box of length $L$ with a spatially uniform distribution of atoms.  Now suppose we turn on a uni-directional barrier somewhere in the box, as shown in Fig.~\ref{fig:2}(a).  After some time, all the atoms will be trapped in one region, as illustrated in Fig.~\ref{fig:2}(b).  

To study this simple model further, we have performed a Monte-Carlo simulation and compared with a simple analytic model.  We start with atoms uniformly distributed in a 1D box and with a Maxwell distribution in velocity with standard deviation $\sigma _v$.  A semi-penetrable wall with width  $2 d$ separates the box into two parts with widths $l_1 > l_2$, so that resonant part of the wall with width $d$ borders with longer side and the blue detuned part of the same width borders the shorter side. We assume that external walls of the box are repulsive for both states. As soon as an atom enters the resonant beam, it gets transferred to state  $\left| 2 \right\rangle $ for which the second half of the wall is attractive~\cite{footnote}. We simulated exponential decay of the atom with decay time $\tau$. As the atom decays it gains one recoil velocity $v _r$ in a random direction. The velocity relaxation time is much longer than time to accumulate in the small region: in the simulation we record velocity of a test particle as soon as it reaches the smaller region and gets a recoil kick in a random direction. Three different cases are considered in the simulation:
(I) Decay occurs in the small region. In this case, the particle is trapped. 
(II) Decay occurs in the large region or in the resonant beam. In this case, the particle is not trapped, but gets another chance and eventually will be trapped.
(III) Decay occurs in the repulsive wall. In that case, the particle is considered lost from the distribution, since it would acquire a large kinetic energy as it exits the barrier.

The model has six parameters:  $d$, $v _r$, $\tau$, $l _1$, $l _2$, $\sigma _v$. The unit of length, $L_u$, is taken to be $d$, and the unit of velocity, $v_u$, is taken to be $v_r$. The unit of time is then $t_u = L_u/v_u = d/v_r$. We observe how a change of parameters ($\tau$, $l_1$, $l_2$, and $\sigma_v$) affects the performance, which we characterize by two figures of merit. The first one is compression in phase space, we define 1D phase space density as the number of trapped particles per unit length per unit velocity,
\begin{equation}
	C = e\frac{{(l_1  + l_2  + 2d) \cdot \sigma _v }}{{l_2  \cdot \sigma _{v,{\rm final}} }},
\end{equation}
where $e$ is the ratio of number of trapped atoms to number of initial atoms. The second figure of merit is the average rate of phase space density change $C / T _f$, where $T _f$ is the time it takes to capture a fraction $f$ of the atoms. For the discussions below we use the time when ninety percent of trappable atoms are captured, $T _{0.90}$.

\begin{figure*}[t]
	\begin{center}
		\includegraphics[width=5.3in]{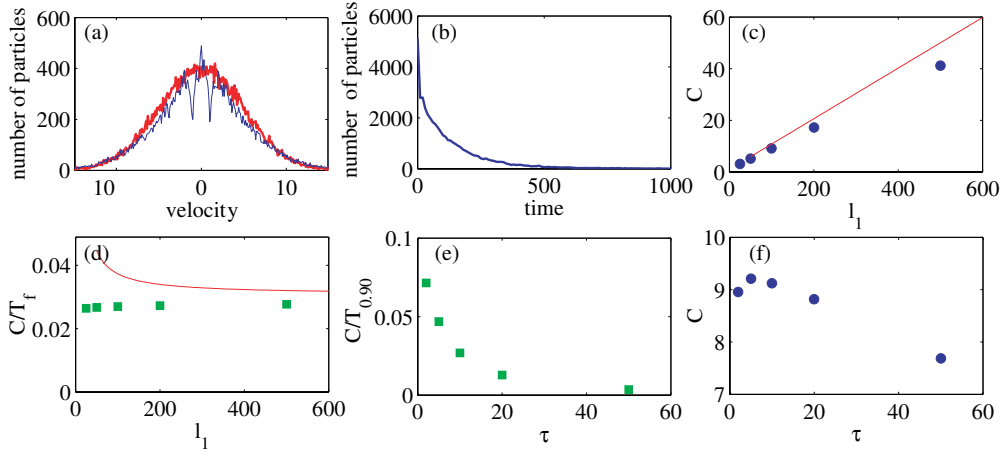}
		\caption{\label{fig:plots-all}(a) Initial and final velocity distributions for parameters  $\tau = 10$, $l_1 = 100$, $l_2 = 10$, $\sigma _v = 5$. Thick line is for initial distribution. Total initial number of particles is 50000. Final distribution is not thermally equilibrated. Dips in it are due to scattering of a single photon. (b) Distribution of times after which particles end up in the smaller region. (c) Change of compression in phase space, solid line is for analytic expression given by analytic formula~(\ref{eq:c-comp}), limiting case~(\ref{eq:c-simp}) is not distinguishable from it in this regime. (d) Average compression rate as size of the larger region $l_1$ is varied, with $f = 0.90$, the lines show the average compression rate estimated from~(\ref{eq:r-simp}) with $f = 0.95$. The numerical solution of~(\ref{eq:non-lin}) give indistinguishable result in this regime as well. (e) and (f) the same when decay time, $\tau$, is varied.}
	\end{center}
\end{figure*}

Figure~\ref{fig:plots-all}(a) shows the velocity distribution for 50000 atoms before and after the process for the following parameters:  $\tau = 10$, $l_1 = 100$, $l_2 = 10$, $\sigma _v = 5$. In the plots (c)-(f) in Fig.~\ref{fig:plots-all}
variations of the parameters are performed with respect to this set. Figure~\ref{fig:plots-all}(b) displays the distribution of capturing times. For this particular set of parameters we find a compression factor, $C = 9.2$.

As the length of bigger part of the box, $l _1$,  increases (Fig.~\ref{fig:plots-all}(c)) the compression factor increases, the average time of the operation increases as well and as a result the rate of compression saturates. For a particular initial velocity distribution and wall width there is an optimal decay time for which the compression is the largest (Fig.~\ref{fig:plots-all}(f)). Average rate of compression in this case decreases monotonically (Fig.~\ref{fig:plots-all}(e)).

Naturally, the operation of the scheme is optimal when the decay rate is much larger than the time most of the particles spend in the gap and much smaller than the time it takes one particle to cross the smaller region: $ t_{\rm gap} \ll \tau \ll t_{\rm travel}$. Also the size of the wall should be much smaller than the size of the both regions: $d \ll l_1 , l_2$. In these limits we can obtain simple analytic expressions for phase space compression and compression rate. 

When we define the fraction of originally trapped particles $f_0 = l_2/(l_1 + l_2)$ the compression in phase space density is given by
\begin{equation}
	C = f_0 \frac{{\sigma _v }}{{\sqrt {f_0 \sigma _v^2  + (1 - f_0 )(\sigma _v^2  + v_r^2 )} }}.
	\label{eq:c-comp}
\end{equation}
In two following limits it becomes
\begin{equation}
	l_1  \gg l_2 ,\sigma _v  \gg v_r ,{\rm ~ ~ ~ ~}C = \frac{{l_1 }}{{l_2 }},
	\label{eq:c-simp}
\end{equation}
\begin{equation}
	l_1  \gg l_2 ,\sigma _r  \ll v_r ,{\rm ~ ~ ~ ~}C = \frac{{l_1 }}{{l_2 }}\frac{{\sigma _v }}{{v_r }},
\end{equation}
hence the scheme is only efficient in the first limit when the initial velocity spread is much larger than the recoil velocity. In this limit it is also applicable in two and three dimensions hence the recoil that might be accumulated in the transverse dimension will not be significant. In Fig.~\ref{fig:plots-all}(c) we show that for appropriate decay times the agreement between this simple analytic formula and the results of Monte-Carlo simulations is very good.

\begin{figure}[b]
\includegraphics[width=2.55in]{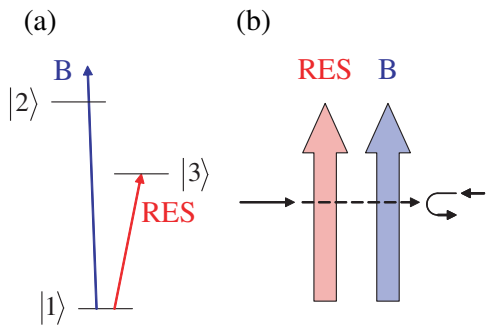}
\caption{\label{fig:3} Extension of the scheme in Fig.~\ref{fig:1} to a three level atom. Transition $\left| 1 \right\rangle  \to \left| 2 \right\rangle $  is a strong dipole transition to create a substantial repulsive wall for state $\left| 1 \right\rangle$. Level $\left| 3 \right\rangle$ is metastable with lifetime comparable to transit time through beams.}
\end{figure}
To estimate the time $T_f$ it takes to capture a fraction $f$ of particles one has to solve the following nonlinear equation
\begin{equation}
	f_0  + (1 - f_0 )\left[ {\frac{1}{{\tilde v_0 }}\sqrt {\frac{2}{\pi }} \left( {1 - e^{ - \frac{{\tilde v_0^2 }}{2}} } \right) + {\rm erfc}\left( {\frac{{\tilde v_0 }}{{\sqrt 2 }}} \right)} \right] - f = 0,
	\label{eq:non-lin}
\end{equation}
here $\tilde v_0 = 2l_1/\sigma_v t$ is velocity, in units of $\sigma_v$, above which all particles are captured in the smaller region. In the limit $l_1 \gg l_2$, $\sigma _v \gg v_r$ and when $\tilde v _0 \ll 1$, i.e. $(1-f) \ll 1$ the equation can be linearized and the average rate is given by
\begin{equation}
	\frac{C}{{T_f }} = \frac{{1 - f}}{{1 - f_0 }}\frac{{\sigma _v }}{{l_2 }}\sqrt {\frac{\pi }{2}} 
	\label{eq:r-simp}
\end{equation}
and becomes independent of $l_1$. Such dependence is seen in Fig.~\ref{fig:plots-all}(d). This simple formula captures the behavior and the result is in reasonable agreement with the simulation, however does not take into account loss of the particles.

As a physical realization of the two-level model we consider a three-level model as illustrated in Fig.~\ref{fig:3}(a).  The ground state $\left| 1 \right\rangle$ has one allowed dipole transition to state $\left| 2 \right\rangle$, and another weak transition to state $\left| 3 \right\rangle$.  Such configuration makes it possible to produce a strong repulsive wall with an allowed dipole transition and a relatively long-lived state for which this wall is nearly transparent. A uni-directional barrier can be constructed in this case in the same way as for the two level model, except that the repulsive barrier should be a beam tuned to the blue of the $\left| 1 \right\rangle  \to \left| 2 \right\rangle $ transition, while the resonant beam is tuned to the $\left| 1 \right\rangle  \to \left| 3 \right\rangle $ transition.  The barrier is illustrated in Fig.~\ref{fig:3}(b).  An atom coming from the right in the ground state is reflected from the repulsive barrier.  
An atom coming from the left first encounters the resonant beam and is excited to the long-lived state $\left| 3 \right\rangle$.  Assuming that the blue-detuned beam is close to the $\left| 1 \right\rangle  \to \left| 2 \right\rangle $  transition, it will generally be completely non-resonant when the atom is in state $\left| 3 \right\rangle$ and the atom can pass through the barrier.

\begin{figure}[t]
\includegraphics[width=2.55in]{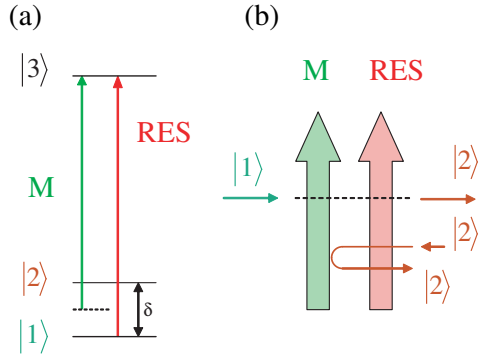}
\caption{\label{fig:4} Scheme that may be used to create a uni-directional wall for the case of alkali atoms. Beam M is attractive for state $\left| 1 \right\rangle$ and repulsive for $\left| 2 \right\rangle$. Beam RES transfers atoms from $\left| 1 \right\rangle$ to $\left| 2 \right\rangle$ in a few scattering events.}
\end{figure}
	This scheme can be realized in alkaline earth atoms.  For example,  calcium has a ground state $\left| g \right\rangle$ and a transition to one excited state $\left| e1 \right\rangle$ with a wavelength of 423 nm and lifetime of 5 ns, and a transition to another excited state $\left| e2 \right\rangle$ with a wavelength of 657 nm and lifetime of 330 $\mu$s.  In this case, the B sheet would be tuned to the blue of the 423 transition (far enough to minimize spontaneous scattering) while the RES sheet would be tuned to the 657 nm transition. The resonant beam must be spectrally broadened in an experimental realization so that Doppler shifts are not important.

	For alkali atoms one-way barrier may be constructed as shown in Fig.~\ref{fig:4}. For atoms in state $\left| 1 \right\rangle$ the beam M is attractive, since it is detuned to the red side of the transition. The state of atoms is changed to $\left| 2 \right\rangle$ by the beam RES in a few scattering events. This state is not affected by the beam RES and the beam M is a repulsive wall for it. For example, in Cesium, which has a ground hyperfine state splitting of 9.2 GHz the beam M would be tuned 4.6 GHz to the red of the $^2 S_{\frac{1}{2}}, {\rm ~}F = 3 \to ^2 P_{\frac{3}{2}},{\rm ~} F = 4$ transition at 852 nm, the RES beam would be tuned to the transition. Phase-space compression in a gravitational trap using a one-way barrier will be analyzed in a separate publication~\cite{future}. In that paper we will also extend the scheme to multilevel atoms and molecules.

	One limitation of the suggested method is that typical dipole trap depths are only a few mK.  This requires therefore in the case of atoms and molecules that cannot be laser-cooled other methods which are not laser-based.  In recent years there 
	has been enormous progress in this direction and several methods have already been demonstrated experimentally.  These include buffer-gas cooling~\cite{doyle98}, Stark deceleration~\cite{bethlem99}, and rotating supersonic nozzle~\cite{gupta01}.

	We can ask about the density limits of the proposed method. Re-absorption of a photon emitted by an atom in the excited state inside the smaller region places the main limitation on density. As the photon mean free path becomes comparable to the average distance an atom travels in the excited state, atoms will leave the region rather than emitted photons and substantial recoil heating will occur. However for the case of the three level model these limits are greatly extended by two factors. The first is the small oscillator strength~\cite{curtis96} and the second is the large Doppler shifts of emitted photon compared with the linewidth. Due to these considerations density will be limited by three-body loss well before photon re-absorption.
	
	In summary, we have shown that atomic phase space compression can be achieved through a variety of related methods  that form a one-way atomic barrier.  As distinct from laser cooling, the methods proposed here do not rely on velocity-selective resonance conditions.  Here, use is instead made of a localized asymmetric excitation in which the order of excitation matters --- so that the structure acts differently on atoms coming from opposite directions thereby encountering excitations in different order.  Interestingly, there is a double analogy between these methods on one hand, and methods to drive current in plasma on the other hand.  Currents can also be driven efficiently in plasma through velocity-selective resonant conditions~\cite{fisch87}, including making use of Doppler shifts, analogous to methods in velocity-selective resonance conditions for laser cooling~\cite{metcalf99,aspect88,kasevich92}.  The one-way wave-induced magnetic wall in the rf regime that drives current in plasma~\cite{fisch03}, does not utilize velocity selection (although, like here, the velocity selection might also be used to additional advantage).  Thus, it is the second rf current-drive effect in plasma, which forms a one-way wall, as opposed to the velocity-selective methods, that is analogous to the effect proposed here of atoms' phase space compression through one-way atomic barriers by asymmetric excitations.   

	The method outlined in this Letter could be used to compress phase space of atoms or molecules that do not have a cycling transition.  It could also be used to initiate evaporative cooling which requires a minimum density to achieve the necessary collision rate.  Finally, a state-selective barrier could find other applications, such as isotope separation.  

	MGR acknowledges support from NSF, the R. A. Welch Foundation, and the S. W. Richardson Foundation and the US Office of Naval Research, Quantum Optics Initiative, Grant N0014-04-1-0336. NJF acknowledges support from the US DOE, under contract DE-AC02-76-CH03073.

\end{document}